\documentclass[aps,preprint]{revtex4}%
\usepackage{amsfonts}
\usepackage{amsmath}
\usepackage{amssymb}
\usepackage{graphicx}%
\begin{document}
\title{Lattice gas models derived from effective field theory}
\author{Matthew Hamilton\footnote{{Current address: Biomathematics Graduate Program,
North Carolina State University, Raleigh, NC 27695}}, Iyam Lynch, Dean Lee}
\affiliation{Department of Physics, North Carolina State University, Raleigh, NC 27695}

\begin{abstract}
We start from a low-energy effective field theory for interacting fermions on
the lattice and expand in the hopping parameter to derive the nearest-neighbor
interactions for a lattice gas model. \ In this model the renormalization of
couplings for different lattice spacings is inherited from the effective field
theory, systematic errors can be estimated a priori, and the breakdown of the
lattice gas model description at low temperatures can be understood
quantitatively. \ We apply the lattice gas method to neutron matter and
compare with results from a recent quantum simulation.

\end{abstract}
\maketitle

\section{Introduction}

Lattice gas models play an important role in many areas of physics from fluid
mechanics \cite{Frisch:1986,Rothman:1988,Gunstensen:1991} to quantum computing
\cite{Meyer:1997,Yepez:1999}, alloy mixing \cite{Muller:1984,Wipf:1996} to
nuclear physics. \ In nuclear physics phenomenological lattice gas models have
been used to model multifragmentation in heavy ion collisions and the
thermodynamics of symmetric and asymmetric nuclear matter at nonzero
temperature
\cite{Pan:1995prc,Pan:1995plb,DasGupta:1996ug,DasGupta:1996ff,Kuo:1996qt,Ray:1997nn,Pan:1998ua,Borg:1998in,Qian:2002kj}%
. \ In this study we attempt to broaden the scope of the lattice gas approach
as applied to interacting fermions at nonzero temperature. \ We build a
connection between lattice gas models and\ low-energy effective field theory
on the lattice. \ While our main interest concerns interacting nucleons, our
approach to lattice gas models should apply to systems such as trapped Fermi
gases near a Feshbach resonance
\cite{Gupta:2002,O'Hara:2002,Gehm:2003,Bourdel:2003,Kinast:2004,Regal:2004},
where similar effective field theory descriptions are applicable.

For most applications of lattice gas models in nuclear physics, the
coefficients of interactions are treated as adjustable parameters tuned to
make the model realistic. \ Also the lattice spacing is often chosen so that
the completely filled lattice corresponds with normal nuclear matter density,
$\rho_{N}\approx0.17$ fm$^{-3}$. \ While one cannot argue with the successes
of this phenomenological approach, there remain some fundamental questions.
\ How do we know which interactions are needed to describe the important
low-energy physics? \ How can we determine the coefficients of the
interactions directly from binding energies and/or few-body scattering data?
\ How can we do simulations at different lattice spacings while keeping the
low-energy physics the same?

In the full quantum theory these questions are answered by effective field
theory. \ In the low-energy limit, power counting schemes organize the
interactions in order of importance
\cite{Weinberg:1990rz,Weinberg:1991um,Kaplan:1996xu,Kaplan:1998tg}. \ Over the
last few years effective field theory methods have been used to study two and
three-nucleon systems at low energy
\cite{Epelbaum:1998na,Beane:2000fx,Bedaque:2002mn} \ There has also been
recent progress in applying effective field theory to many-body nuclear
lattice simulations \cite{Muller:1999cp,Lee:2004si,Lee:2004qd}. \ In this
approach, operator coefficients in the effective lattice Lagrangian are
matched to few-body scattering data, and the renormalization group describes
how the operator coefficients depend on the lattice spacing. \ The resulting
lattice action can then be simulated using standard lattice Monte Carlo
methods to produce many-body results.

In this study we attempt to bring lattice gas models into the framework of
effective field theory. \ We make this connection by means of a spatial
hopping parameter expansion. \ Starting from any effective field theory of
interacting fermions, we show how to construct the rules for a corresponding
lattice gas model. \ We discuss the convergence of the hopping parameter
expansion as well as the uses and limits of the lattice gas approach. \ In
particular we clarify why it can describe a \textquotedblleft
classical\textquotedblright\ phase transition but not a truly
\textquotedblleft quantum\textquotedblright\ phase transition. \ As an example
we consider low-energy neutron matter and compare with recent quantum
simulation results \cite{Lee:2004qd} for the energy per neutron as a function
of density.

\section{Hopping expansion}

Let us consider $n_{f}$ species of fermions and any operator $\hat{V}$ built
up from the annihilation and creation operators. \ Let $\left\vert
\theta\right\rangle $ denote the eigenstates of $\hat{V}$,%
\begin{equation}
\hat{V}(\hat{a}^{\dagger},\hat{a})\left\vert \theta\right\rangle
=V(\theta)\left\vert \theta\right\rangle \text{.}%
\end{equation}
Next we consider a tensor product space that represents the fermions at two
different locations. \ For the moment we single out one particular species,
$j$, and define the Hamiltonian%
\begin{equation}
\hat{H}_{j}=\hat{V}(\hat{a}^{\dagger},\hat{a})\otimes1+1\otimes\hat{V}(\hat
{a}^{\dagger},\hat{a})-\varepsilon\left[  \hat{a}_{j}^{\dagger}\otimes\hat
{a}_{j}+\hat{a}_{j}\otimes\hat{a}_{j}^{\dagger}\right]  .
\end{equation}
Let us define the matrix element%
\begin{equation}
z_{j}^{\theta_{1},\theta_{2}}(\beta)=\left[  \left\langle \theta
_{1}\right\vert \otimes\left\langle \theta_{2}\right\vert \right]  \exp\left[
-\beta\hat{H}_{j}\right]  \left[  \left\vert \theta_{1}\right\rangle
\otimes\left\vert \theta_{2}\right\rangle \right]  \text{.}%
\end{equation}
If we expand in $\beta\varepsilon$ we find
\begin{equation}
z_{j}^{\theta_{1},\theta_{2}}(\beta)=\exp\left[  -\beta(V(\theta_{1}%
)+V(\theta_{2}))\right]  \cdot\left[  1+\beta^{2}\varepsilon^{2}f_{j}%
^{\theta_{1},\theta_{2}}(\beta)\right]  +O((\beta\varepsilon)^{4}%
)\text{,}\label{expand}%
\end{equation}
where
\begin{equation}
f_{j}^{\theta_{1},\theta_{2}}(\beta)=\sum_{\theta^{\prime},\theta
^{\prime\prime}}\left[
\begin{array}
[c]{c}%
|\left\langle \theta^{\prime}\right\vert \hat{a}_{j}\left\vert \theta
_{1}\right\rangle |^{2}|\left\langle \theta^{\prime\prime}\right\vert \hat
{a}_{j}^{\dagger}\left\vert \theta_{2}\right\rangle |^{2}\\
+|\left\langle \theta^{\prime}\right\vert \hat{a}_{j}^{\dagger}\left\vert
\theta_{1}\right\rangle |^{2}|\left\langle \theta^{\prime\prime}\right\vert
\hat{a}_{j}\left\vert \theta_{2}\right\rangle |^{2}%
\end{array}
\right]  \cdot F\left[  \beta(V(\theta^{\prime\prime})+V(\theta^{\prime\prime
})-V(\theta_{1})-V(\theta_{2}))\right]
\end{equation}
and%
\begin{equation}
F(x)=\frac{e^{-x}-1+x}{x^{2}},\qquad F(0)=\frac{1}{2}\text{.}%
\end{equation}

We now generalize this result to a general three-dimensional periodic lattice
Hamiltonian with nearest-neighbor hopping and single-site interactions,%
\begin{equation}
\hat{H}=\sum_{\vec{n}}\hat{V}(\hat{a}^{\dagger}(\vec{n}),\hat{a}(\vec
{n}))-\frac{1}{2m}\sum_{\vec{n}}\sum_{j=1,n_{f}}\sum_{l=1,2,3}\left[  \hat
{a}_{j}^{\dagger}(\vec{n})\hat{a}_{j}(\vec{n}+\hat{l})+\hat{a}_{j}^{\dagger
}(\vec{n})\hat{a}_{j}(\vec{n}-\hat{l})\right]  .
\end{equation}
In the following we use dimensionless lattice parameters. \ If $a$ is the
lattice spacing then the dimensionless mass parameter and inverse temperature
are
\begin{align}
m &  =m_{phys}a\text{,}\\
\beta &  =\frac{1}{T_{phys}a}\text{.}%
\end{align}
Let $\left\vert \Theta\right\rangle $ be a configuration of fermion states at
each lattice site,%
\begin{equation}
\left\vert \Theta\right\rangle =\bigotimes_{\vec{n}}\left\vert \theta(\vec
{n})\right\rangle ,
\end{equation}
Let us define
\begin{equation}
z(\beta,\Theta)=\left\langle \Theta\right\vert e^{-\beta\hat{H}}\left\vert
\Theta\right\rangle .
\end{equation}
After applying the hopping corrections in (\ref{expand}) for each lattice
site, fermion species, and dimension, we find%
\begin{align}
z(\beta,\Theta) &  =\exp\left[  -\beta\sum_{\vec{n}}V(\theta(\vec{n}))\right]
\prod_{\vec{n}}\prod_{j=1,n_{f}}\prod_{l=1,2,3}\left[  1+\left(  \frac{\beta
}{2m}\right)  ^{2}f_{j}^{\theta(\vec{n}+\hat{l}),\theta(\vec{n})}%
(\beta)\right]  \nonumber\\
&  +O((\frac{\beta}{2m})^{4}).\label{main}%
\end{align}
We have assumed that the lattice is longer than three sites in each dimension.
\ If the lattice were three sites long in some dimension then there would be
terms at $O((\frac{\beta}{2m})^{3})$ which wind around the lattice.

We can introduce a chemical potential by adding $-\mu\hat{N}$ to $\hat{H}$,
where $\hat{N}$ is the total fermion number operator. \ In order to compute%
\begin{equation}
z(\beta,\mu,\Theta)=\left\langle \Theta\right\vert e^{-\beta(\hat{H}-\mu
\hat{N})}\left\vert \Theta\right\rangle ,
\end{equation}
we can use the same expression in (\ref{main}) if we redefine%
\begin{equation}
\hat{V}\rightarrow\hat{V}-\mu\sum_{j}\hat{a}_{j}^{\dagger}\hat{a}_{j}\text{.}%
\end{equation}
By summing over all configurations $\Theta$ we now have an approximation to
the grand canonical partition function,%
\begin{align}
Z_{G}  &  =Tr\left[  e^{-\beta(\hat{H}-\mu\hat{N})}\right]  =\sum_{\Theta
}z(\beta,\mu,\Theta)\nonumber\\
&  \approx\sum_{\Theta}\exp\left[  -\beta\sum_{\vec{n}}V(\theta(\vec{n}%
),\mu)\right]  \prod_{\vec{n}}\prod_{j=1,n_{f}}\prod_{l=1,2,3}\left[
1+\left(  \frac{\beta}{2m}\right)  ^{2}f_{j}^{\theta(\vec{n}+\hat{l}%
),\theta(\vec{n})}(\beta)\right]  . \label{grandpart}%
\end{align}
We note the similarity to a $2^{n_{f}}$-state Ising model. \ The only
differences are the non-exponential factors and the temperature dependence in
the interactions. \ One could write everything in exponential form using%
\begin{equation}
1+\left(  \frac{\beta}{2m}\right)  ^{2}f_{j}^{\theta(\vec{n}+\hat{l}%
),\theta(\vec{n})}(\beta)\approx\exp\left[  \left(  \frac{\beta}{2m}\right)
^{2}f_{j}^{\theta(\vec{n}+\hat{l}),\theta(\vec{n})}(\beta)\right]  .
\end{equation}
For weakly coupled systems either form will do. \ However we find that for
very strongly-coupled systems the exponentiated form produces larger
$O((\frac{\beta}{2m})^{4})$ errors than the original expression.

\section{Convergence and long-range order}

The spatial hopping parameter expansion can be extended to higher orders. \ At
order $O((\frac{\beta}{2m})^{n})$ one must consider all $n$-step paths which
are connected and form closed loops. \ On an $L^{3}$ lattice where $L$ is
even, all closed paths must have an even number of steps. \ In that case only
even powers of $\frac{\beta}{2m}$ are nonzero. \ When $L$ is odd, there are
winding paths that give nonzero contributions for odd powers greater than or
equal to $L$. \ A similar expansion was used to derive the zone determinant
expansion in \cite{Lee:2003mb}, where the parameter $\frac{\beta}{2m}$ was
identified as a localization length in lattice units for a given fermion.

As the temperature increases the hopping parameter expansion converges more
quickly. \ However if the temperature is too high then the relevant physics
may be at momenta too high for the chosen lattice spacing. \ The momentum
cutoff scale on the lattice is $\Lambda_{a}=\pi a^{-1}$. \ The temperature
must therefore lie well below the kinetic energy associated with this cutoff
scale,%
\begin{equation}
T_{phys}\ll\frac{\Lambda_{a}^{2}}{2m_{phys}}.
\end{equation}
In order to have a sensible effective theory we need%
\begin{equation}
\frac{1}{\pi^{2}}\ll\frac{\beta}{2m}\text{.}%
\end{equation}
If we combine this with the condition for convergence of the spatial hopping
parameter expansion we get%
\begin{equation}
\frac{1}{\pi^{2}}\ll\frac{\beta}{2m}=\frac{1}{2m_{phys}T_{phys}a^{2}}%
\ll1\text{.} \label{physical}%
\end{equation}

Let us consider as an example $T\approx20$ MeV$,$ roughly the temperature for
the liquid-gas transition in symmetric nuclear matter. \ For a well-defined
lattice gas model based on a hopping parameter expansion, (\ref{physical})
tells us that the lattice spacing must lie in the range from about $1$ fm to
$3$ fm. \ This is enough to probe a wide range of densities, including the
saturation density $\rho_{N}\approx0.17$ fm$^{-3}$. \ Therefore it seems
possible to describe the phase transition at several different lattice spacings.

While a lattice gas model may describe long-range particle density ordering in
a liquid-gas transition, it cannot describe long-range order associated with
\textquotedblleft off-diagonal\textquotedblright\ operators. \ By off-diagonal
operator we mean operators which don't commute with the single-site operator
$\hat{V}$. \ In our lattice gas model formalism these operators are quite
different from diagonal operators, such as the particle density operator,
which commute with $\hat{V}$. \ We can compute the correlation functions
of\ diagonal operators simply by computing the eigenvalues associated with
each configuration state $\left\vert \Theta\right\rangle $. \ But in order to
compute the correlation functions of off-diagonal operators, we need to
consider entirely new hopping paths connecting one operator to another. \ It
is clear that any\ long-range correlations would have to be built by hand from
arbitrarily long paths in our hopping parameter expansion.

Therefore we expect the lattice gas model approach to be ineffective for any
truly \textquotedblleft quantum\textquotedblright\ phase transition.
\ Long-range order in a quantum phase transition becomes possible only when
the quantum wavefunctions of individual particles overlap. \ Hence the
localization length $\frac{\beta}{2m}$ must be greater than the interparticle
spacing in lattice units, and so therefore $\frac{\beta}{2m}\gtrsim1$.

\section{Application to neutron matter}

We now apply our hopping parameter expansion to an effective field theory for
dilute neutron matter on the lattice. \ We focus on low energies and densities
where the relevant momenta are smaller than the pion mass, and we use an
effective field theory with only neutrons. We work with the lowest order
effective Lagrangian which contains a neutron contact interaction that is
adjusted to produce the physical $^{1}S_{0}$ scattering length. \ Our lattice
Hamiltonian with chemical potential included has the form%
\begin{align}
\hat{H}-\mu\hat{N} &  =\left(  m-\mu+\frac{3}{m}\right)  \sum_{\vec{n}}%
\sum_{j=\uparrow,\downarrow}\hat{a}_{j}^{\dagger}(\vec{n})\hat{a}_{j}(\vec
{n})\nonumber\\
&  +C\sum_{\vec{n}}\hat{a}_{\uparrow}^{\dagger}(\vec{n})\hat{a}_{\uparrow
}(\vec{n})\hat{a}_{\downarrow}^{\dagger}(\vec{n})\hat{a}_{\downarrow}(\vec
{n})\nonumber\\
&  -\frac{1}{2m}\sum_{\vec{n}}\sum_{j=\uparrow,\downarrow}\sum_{l=1,2,3}%
\left[  \hat{a}_{j}^{\dagger}(\vec{n})\hat{a}_{j}(\vec{n}+\hat{l})+\hat{a}%
_{j}^{\dagger}(\vec{n})\hat{a}_{j}(\vec{n}-\hat{l})\right]  .\label{neutron}%
\end{align}
We will use the labels $\theta=0,\uparrow,\uparrow,\uparrow\downarrow$ to
represent the various zero, one, and two neutron states on a single site. \ In
Table 1 we list $V(\theta,\mu)$ for the various neutron states.%
\[%
\genfrac{}{}{0pt}{0}{\text{Table 1: \ }V(\theta,\mu)}{%
\begin{tabular}
[c]{|l|l|l|l|}\hline
$0$ & $\uparrow$ & $\downarrow$ & $\uparrow\downarrow$\\\hline
$0$ & $m-\mu+\frac{3}{m}$ & $m-\mu+\frac{3}{m}$ & $2(m-\mu+\frac{3}{m}%
)+C$\\\hline
\end{tabular}
}%
\]
In Table 2 we list $f_{\uparrow}^{\theta_{1},\theta_{2}}(\beta)$ for the
various neutron states on nearest-neighbor sites, and in Table 3 we list
$f_{\downarrow}^{\theta_{1},\theta_{2}}(\beta)$.%
\[%
\genfrac{}{}{0pt}{0}{\text{Table 2: \ }f_{\uparrow}^{\theta_{1},\theta_{2}%
}(\beta)}{%
\begin{tabular}
[c]{|l|l|l|l|l|}\hline
& $0$ & $\uparrow$ & $\downarrow$ & $\uparrow\downarrow$\\\hline
$0$ & $0$ & $\frac{1}{2}$ & $0$ & $F(-\beta C)$\\\hline
$\uparrow$ & $\frac{1}{2}$ & $0$ & $F(\beta C)$ & $0$\\\hline
$\downarrow$ & $0$ & $F(\beta C)$ & $0$ & $\frac{1}{2}$\\\hline
$\uparrow\downarrow$ & $F(-\beta C)$ & $0$ & $\frac{1}{2}$ & $0$\\\hline
\end{tabular}
}%
\]%
\[%
\genfrac{}{}{0pt}{0}{\text{Table 3: \ }f_{\downarrow}^{\theta_{1},\theta_{2}%
}(\beta)}{%
\begin{tabular}
[c]{|l|l|l|l|l|}\hline
& $0$ & $\uparrow$ & $\downarrow$ & $\uparrow\downarrow$\\\hline
$0$ & $0$ & $0$ & $\frac{1}{2}$ & $F(-\beta C)$\\\hline
$\uparrow$ & $0$ & $0$ & $F(\beta C)$ & $\frac{1}{2}$\\\hline
$\downarrow$ & $\frac{1}{2}$ & $F(\beta C)$ & $0$ & $0$\\\hline
$\uparrow\downarrow$ & $F(-\beta C)$ & $\frac{1}{2}$ & $0$ & $0$\\\hline
\end{tabular}
}%
\]

\section{Results}

We have run lattice gas model simulations for both free and interacting
neutron matter. \ The value for the interaction coefficient $C$ is set by
comparing with experimental data from nucleon-nucleon scattering. \ We sum all
nucleon-nucleon scattering bubble diagrams on the lattice, locate the pole in
the scattering amplitude, and compare with L\"{u}scher's formula relating
scattering lengths and energy levels in a finite periodic box
\cite{Luscher:1986pf,Beane:2003da,Lee:2004qd}. \ The results are shown in Table
4.%
\[%
\genfrac{}{}{0pt}{0}{\text{Table 4: Interaction coefficient }C\text{ for
different lattice spacings}}{%
\begin{tabular}
[c]{|l|l|}\hline
$a^{-1}$ (MeV) & $C$ (MeV$^{-2}$)\\\hline
$50$ & $-8.01\times10^{-5}$\\\hline
$60$ & $-6.73\times10^{-5}$\\\hline
$70$ & $-5.81\times10^{-5}$\\\hline
$80$ & $-5.10\times10^{-5}$\\\hline
\end{tabular}
}%
\]
We compute the energy per neutron, $E/A$, as a function of neutron density.
\ The total number of neutrons, $A$, and average energy, $E$, are computed
using
\begin{align}
A &  =\frac{1}{\beta}\frac{\partial}{\partial\mu}\ln Z_{G}\text{,}\\
E &  =-\frac{\partial}{\partial\beta}\ln Z_{G}-(m-\mu)A.
\end{align}

\begin{figure}
[ptb]
\begin{center}
\includegraphics[
height=4.2298in,
width=2.9706in,
angle=-90
]%
{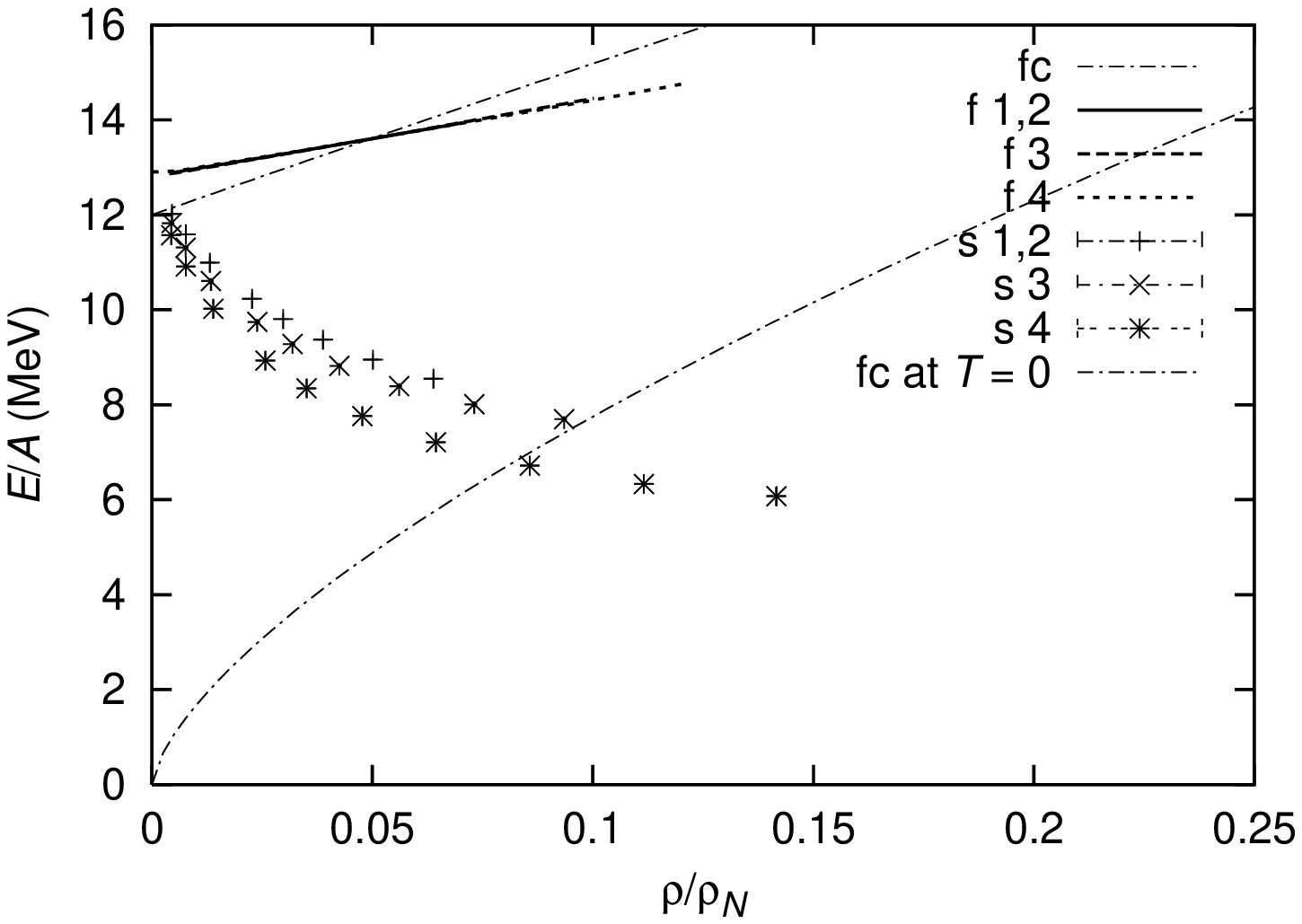}%
\caption{Lattice gas model results for energy per neutron versus density at
$T=8$ MeV.}%
\label{gas_8}%
\end{center}
\end{figure}

\begin{figure}
[ptbptb]
\begin{center}
\includegraphics[
height=4.2298in,
width=2.9706in,
angle=-90
]%
{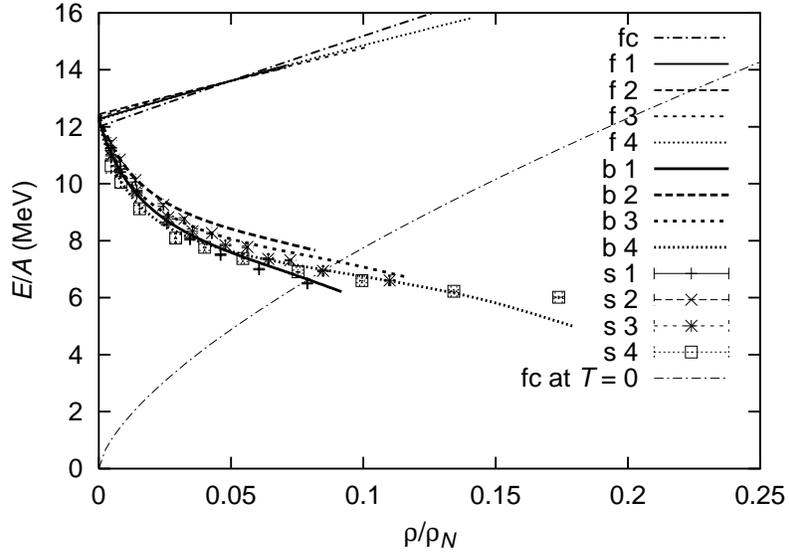}%
\caption{Full quantum effective field theory results for energy per neutron
versus density at $T=8$ MeV.}%
\label{quant_8}%
\end{center}
\end{figure}

The results for $T=8$ MeV are shown in Fig. \ref{gas_8}. \ In Fig.
\ref{quant_8} we show a similar plot from a full quantum lattice simulation
\cite{Lee:2004qd}. \ The lattice volumes for our lattice gas models are chosen
to be the same as that for the corresponding simulations in \cite{Lee:2004qd}.
\ In both plots we use the abbreviation \textquotedblleft fc\textquotedblright%
\ for free continuum results, \textquotedblleft f\textquotedblright\ for free
lattice results, and \textquotedblleft s\textquotedblright\ for lattice
simulation results. \ In Fig. \ref{quant_8}, results for bubble chain diagrams
calculations are also included and labelled with \textquotedblleft
b\textquotedblright\ \cite{Lee:2004qd}. In addition to these abbreviations, we
also use the shorthand labels shown in Table 5 for various combinations of
spatial and temporal lattice spacings.%
\[%
\genfrac{}{}{0pt}{0}{\text{Table 5: Shorthand labels for various lattice
spacings}}{%
\begin{tabular}
[c]{|l|l|l|}\hline
$a^{-1}($MeV$)$ & $a_{t}^{-1}($MeV$)$ & Label\\\hline
$50$ & $24$ & $0$\\\hline
$60$ & $32$ & $1$\\\hline
$60$ & $48$ & $2$\\\hline
$70$ & $64$ & $3$\\\hline
$80$ & $72$ & $4$\\\hline
\end{tabular}
\ \ }%
\]
\ For the lattice gas model however the temporal lattice spacing, $a_{t}$, is
set to zero.

\begin{figure}
[ptb]
\begin{center}
\includegraphics[
height=4.2298in,
width=2.9706in,
angle=-90
]%
{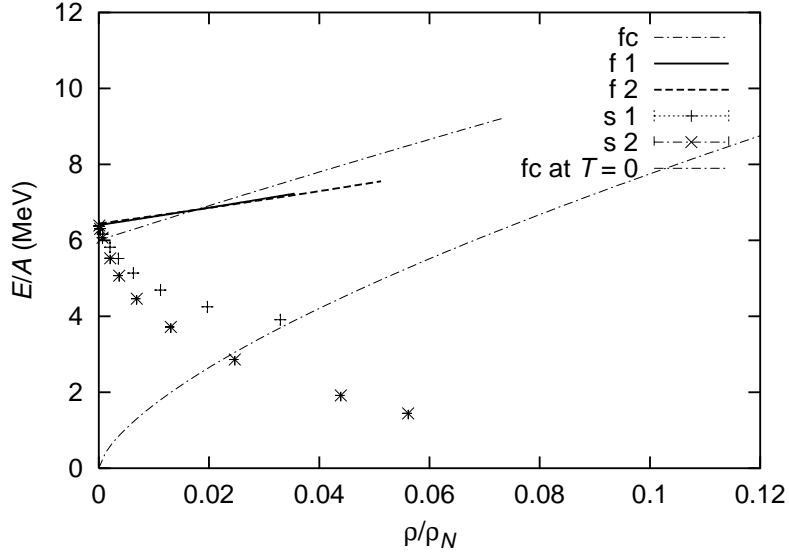}%
\caption{Lattice gas model results for energy per neutron versus density at
$T=4$ MeV.}%
\label{gas_4}%
\end{center}
\end{figure}

\begin{figure}
[ptbptb]
\begin{center}
\includegraphics[
height=4.2298in,
width=2.9706in,
angle=-90
]%
{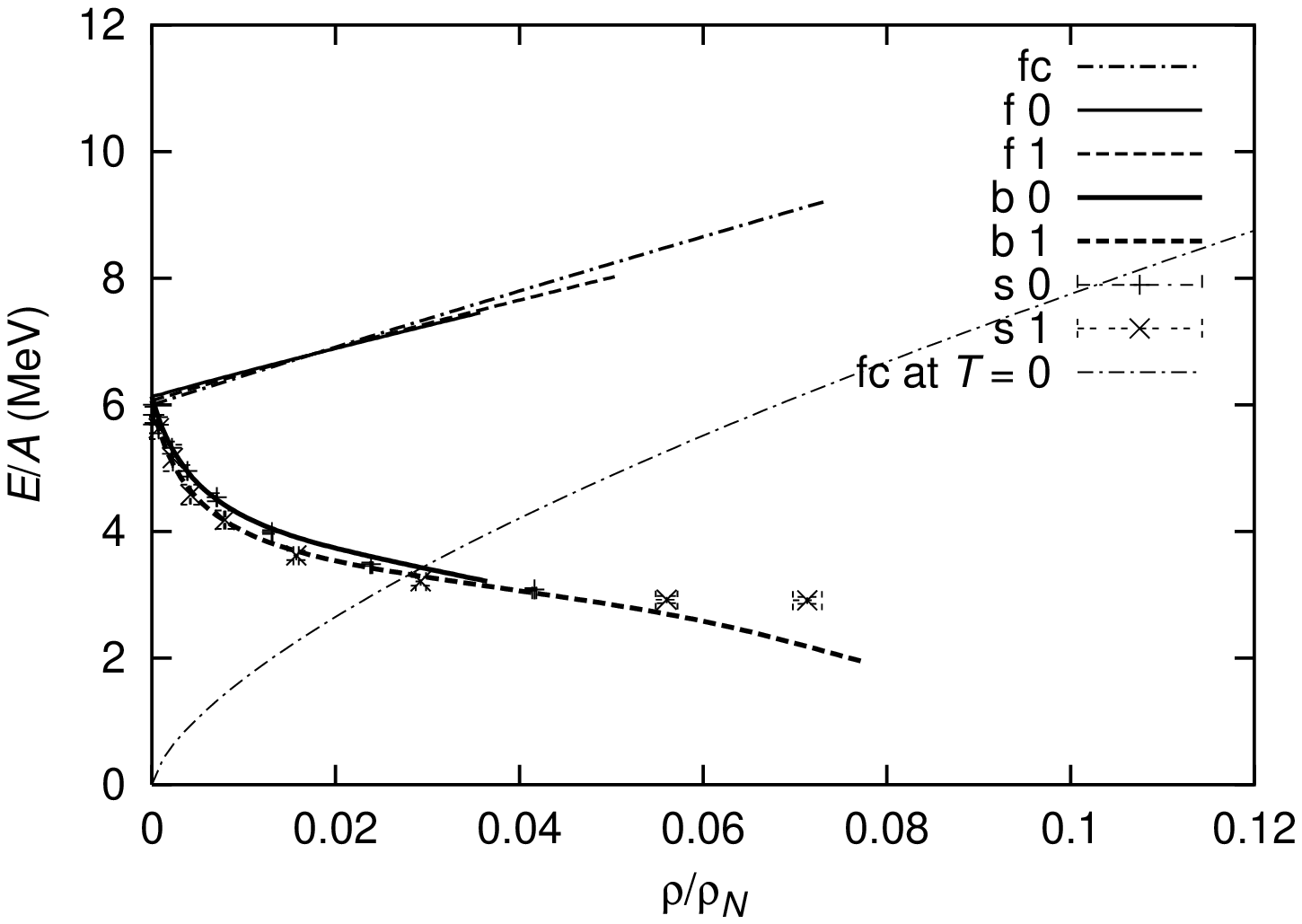}%
\caption{Full quantum effective field theory results for energy per neutron
versus density at $T=4$ MeV.}%
\label{quant_4}%
\end{center}
\end{figure}

In Fig. \ref{gas_4} we show results at $T=4$ MeV for the lattice gas model,
and in Fig. \ref{quant_4} we show the full quantum simulation at $T=4$ MeV.%
For each of the temperatures and spatial lattice spacings we show the
corresponding spatial hopping parameter\ in Table 6.%
\[%
\genfrac{}{}{0pt}{0}{\text{Table 6: \ Hopping parameters for various lattice
spacings}}{%
\begin{tabular}
[c]{|l|l|l|l|l|}\hline
$a^{-1}($MeV$)$ & $50$ & $60$ & $70$ & $80$\\\hline
$\frac{\beta}{2m}$ for $T=8\text{ MeV}$ & $0.17$ & $0.24$ & $0.33$ &
$0.43$\\\hline
$\frac{\beta}{2m}$ for $T=4\text{ MeV}$ & $0.33$ & $0.48$ & $0.65$ &
$0.85$\\\hline
\end{tabular}
}%
\]

As discussed in the previous section, the lattice gas model cannot describe
the superfluid transition in neutron matter since this requires long-range
ordering associated with neutron pairing. \ There may be some indication of
this already in the $T=4$ MeV data shown in Fig. \ref{gas_4}. \ The points at
higher density show considerable deviation from the full quantum simulation
results in Fig. \ref{quant_4}. \ Nevertheless we see that the lattice gas
results agree quite well with the quantum simulations when the spatial hopping
parameter is less than about $0.4$. \ This is a bit surprising considering
that the computational cost for the lattice gas model simulation is several
hundred times less than the quantum simulation.

\section{Summary}

Starting from a low-energy effective field theory for interacting fermions on
the lattice, we derive the nearest-neighbor interactions for a lattice gas
model by expanding in the spatial hopping parameter. \ Unlike most
phenomenological approaches, we derive equivalent lattice gas models at
different lattice spacings and determine coefficients directly from binding
energies and/or few-body scattering data. \ We also give an estimate of the
systematic errors and discuss the limits of the lattice model approach in
describing long-range ordering. \ As a concrete example we apply the effective
field theory lattice gas approach to low-energy neutron matter and compare
with results from a recent quantum simulation. \ Despite the very low
computational cost, essentially the same as that for a $2^{n_{f}}$-state 3D
Ising model, we find good agreement with full quantum simulation results when
the hopping parameter is not too large.

In our approach temperature-dependent interactions are naturally introduced into the lattice gas model. \ These are necessary to reproduce the physics of the full quantum theory at different temperatures.  To our knowledge, neutron matter at this density and temperature has never
been previously described using a lattice gas model. \ We hope that the
broader application of lattice gas models as well as the connection with
effective field theory will prove useful in the study of other many-body
fermion systems at nonzero temperature.\medskip

This work was supported in part by the U.S. Department of Energy under grant
and DE-FG02-04ER41335 and the Minority Graduate Education Program at North
Carolina State University.

\bibliographystyle{h-physrev3}
\bibliography{NuclearMatter}

\end{document}